\documentstyle[aps, prb, epsf, multicol]{revtex}
\begin{document}
\title{Exact ground state of the generalized three-dimensional
Shastry-Sutherland model} 
   
\author{Shu Chen and H. B\"{u}ttner} 
\address{ 
Theoretische Physik 1, Universit\"{a}t Bayreuth, D-95440 Bayreuth, Germany}
\date{\today}
\maketitle

\begin{abstract}
We generalize the Shastry-Sutherland model to three dimensions. By
representing the model as a sum of the semidefinite positive projection operators,
we exactly prove that the model has exact dimer ground state. Several
schemes for constructing the three-dimensional Shastry-Sutherland model are proposed.\\ 


\end{abstract}

\begin{multicols}{2}
\narrowtext

There has been an increasing interest in the Shastry-Sutherland (S-S)
model\cite{S-S} since it can describe many aspects of the two-dimensional spin
gap system $SrCu_2(BO_3)_2$.\cite{Kageyama,Miyahara} In a series of theoretical
investigations, various aspects of the S-S model have been described.
\cite{Chung,Zheng,Balents,Totsuka,Misguich} 

The S-S model is a two dimensional square lattice antiferromagnet with
additional diagonal interactions in every second square with alternating
directions, see Fig. 1. For the square lattice interaction $J$ and the diagonal
interaction  $J_d$, the Hamiltonian can be written as   
\begin{equation}
H = J  \sum_{\langle i,j \rangle}  {\bf S}_i  \cdot {\bf S}_{j} + 
    J_d  \sum_{\langle i,j \rangle_d}  {\bf S}_{i}  \cdot {\bf S}_{j}
\label{ss}
\end{equation}
Shastry and Sutherland have shown that the product of singlet pairs (dimers)
along the diagonal bonds is the ground state of the system for $ J_d
\geq 2 J $.     
\begin{figure}
\centerline{\epsfysize=5.cm \epsffile{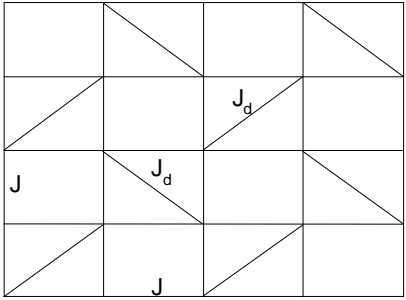}}  
\caption{The Sharstry-Sutherland lattice.}   
\label{fig1}  
\end{figure}

\begin{figure}
\centerline{\epsfysize=5.cm \epsffile{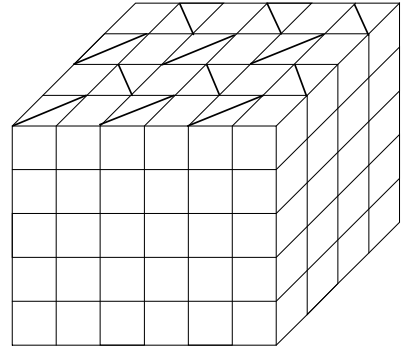}}  
\caption{The generalized 3D Sharstry-Sutherland model.}   
\label{3d}  
\end{figure}

\begin{figure}
\centerline{\epsfysize=5.cm \epsffile{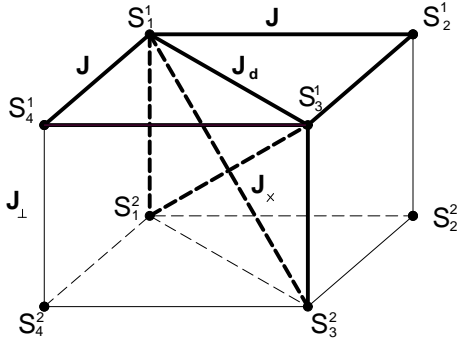}}  
\caption{The basic cubic unit with inlayer and interlayer diagonal exchange
interactions which constructs the 3D Sharstry-Sutherland model. (See text for
further discussion)}     
\label{cubic}  
\end{figure}

In this paper we generalize the S-S model to three dimensions with
various types of interactions. Between the different planes we can construct
the exact ground states of three-dimensional models and by this increase the
number of three-dimensional models with the exactly known ground states. We
consider a three-dimensional cubic lattice (Fig.\ref{3d}) constructed from
basic cubic units shown in Fig.\ref{cubic}. Each layer in this cubic system
is a S-S lattice which is coupled to the next layer by the perpendicular
interlayer interaction $J_{\perp}$ and the diagonal interaction
$J_{\times}$ connected the diagonal end points to those in the next layer.
That means that the Sharstry-Sutherland diagonals are on top of each other.
If we label the different layers by $\alpha = 1, \cdots, L$ and have $N \times
M$ sites per layer, the Hamiltonian can be written as       
\begin{eqnarray}      
H & = & J   \sum_{\langle i,j \rangle, \alpha}  {\bf
S}_{i}^{\alpha} \cdot {\bf S}_{j}^{\alpha} +  J_d  \sum_{\langle i,j
\rangle_d, \alpha }  {\bf S}_{i}^{\alpha}  \cdot {\bf S}_{j}^{\alpha}
\nonumber \\  & + &  J_{\times}  \sum_{\langle i,j \rangle_{d'}, \alpha}
\left( {\bf S}_{i}^{\alpha}  \cdot {\bf S}_{j}^{\alpha+1} + {\bf
S}_{j}^{\alpha}  \cdot {\bf S}_{i}^{\alpha+1} \right) \nonumber\\ &+&
J_{\perp} \sum_{ i , \alpha}   {\bf S}_{i}^{\alpha} \cdot {\bf
S}_{i}^{\alpha+1}\; ,   \label{3dss}      
\end{eqnarray}  
Periodic boundaries in each layer are
assumed and $M$ and $N$ have to be even, while $L$ can be even
or odd.   

\begin{figure}
\centerline{\epsfysize=5.cm \epsffile{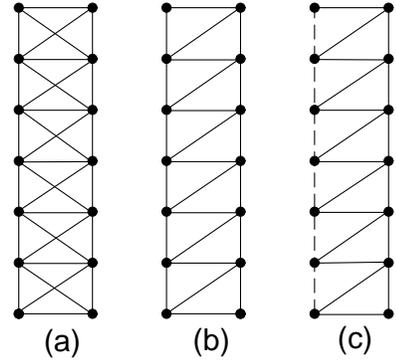}}  
\caption{The spin ladders correspond to the 3D S-S model with $J=0$.}    
\label{ladd}  
\end{figure}

It is instructive to consider two different parameter limits of our model
(\ref{3dss}): (i) For $J_{\perp} = J_{\times} = 0 $,  the 3D model reduces to
$L$ independent S-S layers. (ii) For $J = 0 $ within each layer, the
three-dimensional system decouples to independent $\frac{1}{2} N \times M$ 
two-leg spin ladders of length $L$. The structure of this ladder is shown in
Fig.\ref{ladd} (a). It is known that such a spin ladder with $J_{\perp} =
J_{\times}$ has exact ground state composed of a product of dimers along the
rungs of ladder \cite{Bose} as long as the exchange interaction along the rung
satisfies the condition $J_d \geq 2 J_{\perp}$. This means that in both of the
special cases, the product of singlet pairs along the inlayer diagonal bonds
is the exact ground state of the model (\ref{3dss}). In the following,  we
will prove that for the general case (but $J_{\perp} = J_{\times}$) the ground
state of the three-dimensional model is given by the product of all diagonal
singlet pairs   
\begin{equation}  
\label{layer-dimer}  
\Phi _{D} = \prod_{\alpha=1}^{L} \prod_{\langle i,j \rangle_d}
\frac{1}{\sqrt{2}} \left( [\uparrow]_{i}^{\alpha} [\downarrow]_{j}^{\alpha}
-[\downarrow]_{i}^{\alpha} [\uparrow]_{j}^{\alpha} \right)   
\label{product}
\end{equation}
for special condition 
\begin{equation} 
 J_{d}  \geq  2(J + J_{\perp}) 
\label{constraint}
\end{equation} 

The rigorous proof is made by representing the above model as
a sum of the positive semidefinite projection operators and the above condition
(\ref{constraint}) guarantees that $\Phi_{D}$ is the ground state of the
system. The proof furthermore employs the fact that the global Hamiltonian
(\ref{3dss}) can be written as a sum of many local sub-Hamiltonians defined on
the basic cubic unit (see Fig.\ref{cubic}). There are altogether $Q =
\frac{1}{2} N \times M \times L $ units as local sub-Hamiltonians $h$. This
part of the Hamiltonian can be written as
\begin{eqnarray}
h &=& J_d  {\bf S}_{1}^{1} \cdot {\bf S}_{3}^{1} + J ({\bf S}_{1}^{1} + {\bf
S}_{3}^{1}) \cdot ({\bf S}_{2}^{1} + {\bf S}_{4}^{1}) \nonumber \\ 
&+& J_{\perp} ({\bf
S}_{1}^{1} +  {\bf S}_{3}^{1}) \cdot  ({\bf S}_{1}^{2} + {\bf S}_{3}^{2}) ,  
\end{eqnarray} 
which is just the sum of spin exchange interactions along the bonds
represented by bold lines in Fig. \ref{cubic}. Those spin interactions
represented by thin lines belong to the neighboring cubes.

We now define a projection operator ${\bf P}$ composed of three one half spins
as  
\begin{equation}  
{\bf P} [{\bf S}_{1}, {\bf S}_{2}, {\bf S}_{3}] = \frac{1}{3} \left[
\left({\bf S}_{1}+ {\bf S}_{2}+ {\bf S}_{3}  \right)^{2} - \frac{3}{4}
\right],    
\end{equation} 
which projects a state into the subspace with total spin $3/2$.    
Using the projection operators, we can transform our Hamiltonian part $h$ which
is represented as 
\begin{eqnarray}   
h &=& - \frac{3}{4} (2J + 2J_{\perp}) +
  (J_d -2 J - 2 J_{\perp})  {\bf S}_{1}^{1} \cdot {\bf S}_{3}^{1} \nonumber \\
  & & + J~~ \left[ {\bf P}({\bf S}^{1}_{1}, {\bf S}^{1}_{2}, {\bf
S}^{1}_{3})  + {\bf P}({\bf S}^{1}_{1}, {\bf S}^{1}_{4}, {\bf S}^{1}_{3})
\right]  \nonumber \\
  & & + J_{\perp} \left[ {\bf P}({\bf S}^{1}_{3}, {\bf S}^{1}_{1}, {\bf
S}^{2}_{1})  + {\bf P}({\bf S}^{1}_{1}, {\bf S}^{1}_{3}, {\bf S}^{2}_{3})
\right] 
\; .  
\label{h}      
\end{eqnarray}  
It is obvious that, for $J_d = 2(J + J_{\perp})$, certain terms in Eq.
(\ref{h}) vanish and thereby the Hamiltonian is a sum of four positive
semidefinite projection operators. The singlet state
\begin{equation}
\left[ {\bf S}^{1}_{1}, {\bf
S}^{1}_{3}\right] = \left( [\uparrow]_{1}^{1} [\downarrow]_{3}^{1}
-[\downarrow]_{1}^{1} [\uparrow]_{3}^{1} \right)/ \sqrt{2} 
\end{equation} 
has the lowest eigenvalue $0$ for each of the four projection operators
and thus is the ground state of this sub-Hamiltonian. For larger $J_d $ this
singlet is also the lowest energy eigenstate of the term ${\bf S}_{1}^{1}
\cdot {\bf S}_{3}^{1}$, and hence it is the ground state of the total
sub-Hamiltonian $h$ with the ground state energy $E_h = - \frac{3}{4} J_d$.
All the other sub-Hamiltonians defined on other basic units of cube have the
same properties as the one explicitly shown in Fig. \ref{cubic}. Therefore,
the global ground state of this three-dimensional model is just a product of
dimers for each layer. Such a ground state is essentially an {\em optimum
ground state} of the global Hamiltonian, since it is simultaneously ground
state of every local sub-Hamiltonian.\cite{Niggemann} The corresponding ground
state energy is then given by
\begin{equation} 
E = - \frac{3}{4} Q J_d .
\end{equation}

This general proof actually does not depend on the special coupling parameters
between the layers we discussed so far. We see that the main condition is that
the dimer along the diagonals in the layer should be also the dimers along the
rungs of the corresponding ladders. And as long as the vertical spin ladders
have dimers along the rungs as ground states, we can put them together in
three dimensions. Therefore, generalization of the model (\ref{3dss}) is
straightforward by changing the inter-layer coupling way. The first example is
that we couple the different layers only along one interlayer diagonal
(Fig.\ref{ladd}. (b)). In this case we require $J_{\times} = 2 J_{\perp}$
which means that the individual ladders are Majumdar-Gosh chains.\cite{MG} The
second example is given when the two legs of ladder have different interactions
(Fig. \ref{ladd}. (c)).\cite{CBV,Chen} If we call them  $J_{\perp}$ and
$J'_{\perp}$, we should have the conditions 
\begin{equation}
J_{\times} = J_{\perp} +
J'_{\perp}
\end{equation}
and
\begin{equation} 
J_d \geq 2J + J_{\perp} + J'_{\perp} .
\end{equation} 
Even the limit $J'_{\perp}=0$ is allowed and we have vertical sawtooth
chains coupled together. \cite{Sen,Nakamura} In the third example, we may
couple the different layers by some intermediate spins shown in Fig.5. If this
coupling has a strength $J'$, we should have the condition
\begin{equation} 
J_d \geq 2J + J' 
\end{equation} 
and the ground state is still a product of dimers each layer. There could be
of course other constructions. The main point is that we have shown that the
Shastry-Sutherland layer could be coupled to other layers by different
mechanisms which only have to make sure that these ladders have dimers on the
rungs as ground state.
\begin{figure} 
\centerline{\epsfysize=3.cm
\epsffile{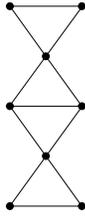}}   
\caption{The dimers on neighboring layers (rungs) have
no direct coupling, but couple to an intermediate spin between them with
exchange strength of $J'$.}   
\label{f5}   
\end{figure}

We have limited our discussion to homogeneous models in which
every basic unit has the same structure and exchange strength. We
can also generalize our discussion to inhomogeneous cases, e.g. to a model
where the inlayer coupling strengths, $J^{\alpha}$ and
$J^{\alpha}_d$, are different in each layer. In this case we only have to make
sure that the constraint
\begin{equation}
J^{\alpha}_d \geq 2(J^{\alpha} + J_{\perp})
\end{equation}
is still fulfilled.  

To conclude, we extend the two dimensional S-S model to three dimension and
find the exact ground state of the generalized three-dimensional S-S model.
An exact proof based upon the representation of projection operators is given.
We also discuss several ways of extending the 3D S-S model.  

S. Chen would like to specially thank J. Voit for his critical reading of the
manuscript and encouragement. This research was supported by DFG through Grants
No. VO436/7-2. 

\end{multicols}
\end{document}